\def\H        {{$^1$H \/}}

\def\C        {{$^{13}$C \/}}

\def\NN       {{$^{15}$N \/}}

\newcommand{\mr}[1]{\mathrm{#1}}
\newcommand{\unit}[1]{\,\mathrm{#1}}

\newcommand{\us}{\,\mu{\rm s}}

\newcommand{\Sx}{S_x}

\newcommand{\Sz}{S_z}
\newcommand{\Se}{S_e}
\newcommand{\Ix}{I_x}

\newcommand{\Iz}{I_z}
\newcommand{\Ie}{I_e}
\newcommand{\Ucp}{U_\mr{CP}}
\newcommand{\iUcp}{\Ucp^\dagger}
\newcommand{\Ufree}{U_\mr{free}}
\newcommand{\iUfree}{\Ufree^\dagger}
\newcommand{\Hfree}{H_\mr{free}}

\newcommand{\ma}{|0\rangle}
\newcommand{\ima}{\langle 0|}
\newcommand{\mb}{|1\rangle}

\newcommand{\sa}{|a\rangle}
\newcommand{\isa}{\langle a|}

\newcommand{\apar}{a_{||}}
\newcommand{\aperp}{a_\bot}

\newcommand{\won}{\omega_\mr{0,n}}

\newcommand{\yn}{\gamma_\mr{n}}
\newcommand{\tr}{\mr{trace}}

\newcommand{\captionstyle}{\normalfont} %define a caption font size

\documentclass[prl,aps,twocolumn]{revtex4-1}% PRL

\usepackage{graphicx}
\usepackage{bm}% bold math
\usepackage{amsmath}
\usepackage{amssymb}
\usepackage{verbatim}
\usepackage[colorlinks=true, pdfstartview=FitV, linkcolor=blue, citecolor=blue, urlcolor=blue]{hyperref} % enable links

\begin{document}

% formatting
\global\emergencystretch = .1\hsize % adjust the line-breaking to avoid overfull \hbox (standard .15\hsize)

%\title{One- and two-dimensional NMR spectroscopy with a diamond quantum sensor}
\title{One- and two-dimensional nuclear magnetic resonance spectroscopy\\ with a diamond quantum sensor}

\author{J. M. Boss$^1$, K. Chang$^1$, J. Armijo$^2$, K. Cujia$^1$, T. Rosskopf$^1$, J. R. Maze$^2$, and C. L. Degen$^1$}
  \email{degenc@ethz.ch} 
  \affiliation{
   $^1$Department of Physics, ETH Zurich, Otto Stern Weg 1, 8093 Zurich, Switzerland.
	 $^2$Departmento de F\'isica, Pontificia Universidad Cat\'olica de Chile, Santiago 7820436, Chile.
	}
	
%\date{\today}

\begin{abstract}
We report on Fourier spectroscopy experiments performed with near-surface nitrogen-vacancy centers in a diamond chip.  By detecting the free precession of nuclear spins rather than applying a multipulse quantum sensing protocol, we are able to unambiguously identify the NMR species devoid of harmonics.  We further show that by engineering different Hamiltonians during free precession, the hyperfine coupling parameters as well as the nuclear Larmor frequency can be selectively measured with high precision (here 5 digits).  The protocols can be combined to demonstrate two-dimensional Fourier spectroscopy.  The technique will be useful for mapping nuclear coordinates in molecules en route to imaging their atomic structure.
\end{abstract}

\pacs{76.70.Hb, 76.30.Mi, 03.65.Ta, 82.56.-b}
%76.70.Hb 	Optically detected magnetic resonance (ODMR)
%76.30.Mi 	Color centers and other defects
%03.65.Ta 	Foundations of quantum mechanics; measurement theory
%82.56.Dj 	High resolution NMR
%82.56.Fk 	Multidimensional NMR
%82.56.Jn 	Pulse sequences in NMR
%82.56.-b 	Nuclear magnetic resonance

\maketitle

%\tableofcontents

%%%%%%%%%%%%%% Introduction

Nitrogen-vacancy (NV) centers in diamond have opened exciting perspectives for the ultrasensitive detection of nuclear magnetic resonance (NMR), with possible applications to molecular structure imaging and chemical nanoanalytics \cite{degen08apl,mamin13,staudacher13}.  NMR signals are detected by placing an analyte on a diamond chip engineered with a surface layer of NV centers, and measuring the weak magnetic dipole fields of nuclei via optically detected magnetic resonance \cite{davies76,gruber97}.
Examples of the rapid recent progress in NV-NMR include the detection of small numbers of nuclei within voxels of a few (nm)$^3$ \cite{loretz14apl,muller14}, the detection of multiple nuclear isotopes \cite{devience15,haberle15} and naturally occurring adsorption layers \cite{loretz14apl}, the observation of surface diffusion and molecular motion \cite{kong15,staudacher15}, scanning imaging with $<20\unit{nm}$ spatial resolution \cite{haberle15,rugar15}, and the spatial mapping of up to 8 internal \C nuclei \cite{schirhagl14}.  One of the far goals of NV-NMR is the detection and three-dimensional localization of individual nuclei in single molecules deterministically placed on the diamond chip \cite{degen08apl,zhao11,schirhagl14}.

\begin{figure}[t!]
\includegraphics[width=1.00\columnwidth]{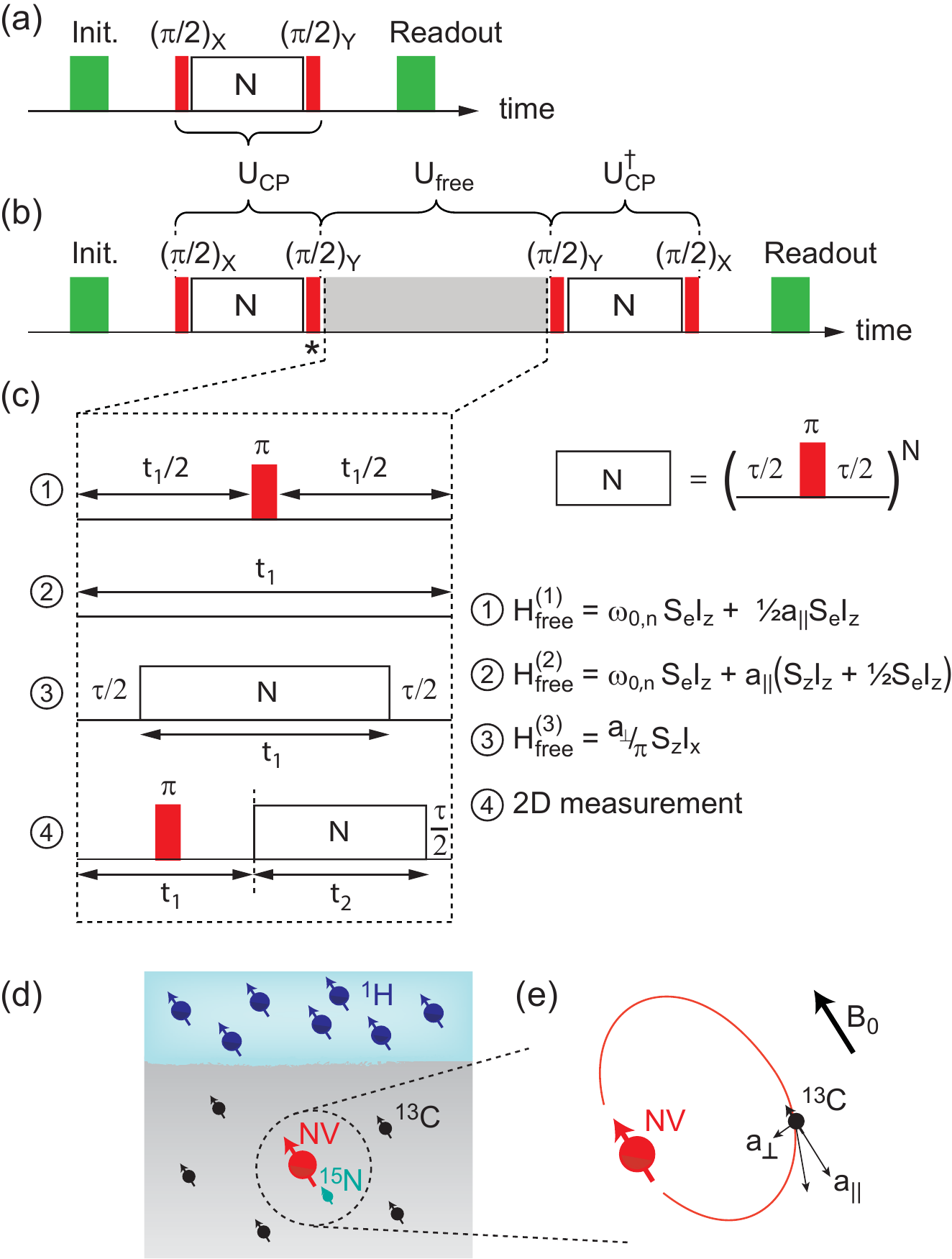}
\caption{\captionstyle
(a) Pulse protocol for multipulse spectroscopy.  Laser pulses are colored in green and microwave pulses are colored in red.  The $\pi$ pulse series used the XY8 phase cycling pattern of Ref. \cite{gullion90}.
(b) Pulse protocol for a free nuclear precession measurement.  Pulse marked by $\ast$ was used for phase cycling (phases $\mr{Y}$, $\mr{\bar{Y}}$).
(c) Different free-precession sequences and corresponding Hamiltonians $\Hfree^{(1...3)}$ exploited in our experiments.  Hamiltonians are given in units of angular frequency and only the relevant secular part is shown.  $t_1$ and $t_2$ are incremented evolution periods.
(d) Sketch of diamond chip with a near-surface NV center and \NN, \C and \H nuclei.
(e) Vectorial picture of the parallel and transverse hyperfine parameters, $\apar$ and $\aperp$ (see Refs. \cite{taminiau12,loretz14science} for a definition).  Contour represents a field line of the NV magnetic dipole.
}
\label{fig:fig1}
\end{figure}

Sensitive detection of nuclear magnetic signals is possible with multipulse sequences that consist of a series of $\pi$ pulses (see Fig. \ref{fig:fig1}a).  These sequences act like a narrow-band lock-in amplifier \cite{kotler11} whose demodulation frequency $f=1/(2\tau)$ is set by the delay time $\tau$ between the pulses \cite{cywinski08,zhao11,delange11}.  By varying $\tau$ a frequency spectrum of the magnetic field can be recorded.  Multipulse spectroscopy of NMR signals has been reported for many nuclear isotopes, including $^1$H, $^{13}$C, $^{14}$N, $^{15}$N, $^{19}$F, and possibly $^{29}$Si and $^{31}$P \cite{staudacher13,loretz14apl,muller14,loretz15,rugar15,haberle15,devience15}.  These experiments have, however, also revealed some important shortcomings of the method, including a modest spectral resolution \cite{mamin13,laraoui13} and ambiguities in peak assignments due to signal harmonics \cite{loretz15}. The fundamental reason for both effects is the indirect way nuclear spin signals are detected via their influence on the electronic spin.

A more natural way for measuring NMR signals is to observe the free nuclear precession in the absence of microwave or radio-frequency pulses, reminiscent of the ``free induction decay'' in conventional NMR Fourier spectroscopy.  The free nuclear precession can be detected by performing two consecutive nuclear spin measurements and incrementing the duration $t_1$ between the measurements.  Mamin et al. \cite{mamin13} have demonstrated Fourier NMR by an electron-nuclear double resonance technique.  Laraoui et al. \cite{laraoui13} and others \cite{kong15,staudacher15} have used two successive multipulse sequences to directly record the nuclear precession.

In the present work, we demonstrate two important applications of the free nuclear precession method: The unambiguous assignment of NMR peaks to nuclear isotopes, and a set of measurements for determining nuclear hyperfine couplings with high precision.  By combining two free precession periods, we further demonstrate two-dimensional Fourier spectroscopy \cite{aue76} of nuclear spins.  The presented techniques will be useful for mapping nuclear coordinates in more complex molecular structures.

The basic sequence of a free nuclear precession experiment is shown in Fig. \ref{fig:fig1}b.  The sequence consists of three propagators, 
\begin{equation}
\Ucp^\dagger \Ufree \Ucp \ ,
\label{eq:rotations}
\end{equation}
where $\Ucp$ represents the traditional multipulse sequence \cite{delange11} and $\Ufree = e^{-i\Hfree t_1}$ describes the period of free precession.  $\Hfree$ is the free precession Hamiltonian that is varied in our experiments.  (Note that we use units of angular frequency).
The measured ``signal'' is the probability $p$ that the initial quantum state $\ma$ of the electronic sensor spin is recovered after the three propagations,
\begin{equation}
p = |\ima\Ucp^\dagger \Ufree \Ucp \ma|^2 = |\isa \Ufree \sa|^2 \ ,
\label{eq:p}
\end{equation}
where $\sa = \Ucp \ma$ is the state after the $\Ucp$ rotation.
%For NV centers, initialization and readout of the spin state $\ma$ is possible using laser pulses.

Eq. (\ref{eq:p}) contains the essence of the free precession measurement.  While conventional multipulse spectroscopy detects the electronic evolution $\Ucp$, with the signal given by $p = |\ima\Ucp\ma|^2$, the free precession sequence measures the nuclear evolution $\Ufree$ and simply applies $\Ucp$ and $\iUcp$ to transform from and to the detectable electronic state $\ma$.

We now walk through the implementation of the three unitary rotations in some detail (see Fig. \ref{fig:fig1}).
For simplicity, we focus on the detection of a single nuclear spin $I=1/2$.  The basic free Hamiltonian in the absence of any pulses is
\begin{equation}
\Hfree = \won \Iz + \apar \Sz\Iz + \aperp \Sz\Ix  \ ,
\end{equation}
where $\apar$ and $\aperp$ are the parallel and transverse hyperfine coupling parameters (see Fig. \ref{fig:fig1}e) \cite{taminiau12,loretz14science}, $\won = \yn B_0 \gg \aperp$ is the bare nuclear Zeeman frequency, $\yn$ is the nuclear gyromagnetic ratio, $B_0$ is the magnetic bias field, $S_k$, $I_k$ are $\frac{1}{2}\times$ the Pauli matrices ($k=x,y,z$), and where we are in the electron's rotating frame of reference.  (To include multiple nuclear spins, the Hamiltonian can simply be summed over the contributions of individual nuclei).

For the following, it is convenient to represent the quantum state of the coupled spin system by its density matrix $\sigma$.  After initialization into $\ma$, the density matrix is $\sigma_0 = \ma\ima\otimes\Ie = (\Sz + \Se)\otimes\Ie$, where $\Se$ and $\Ie$ are $\frac{1}{2}\times$ the identity matrix.  For the sake of brevity we will in the following omit $\Se$ and $\Ie$, such that
\begin{equation}
\sigma_0 = \ma\ima = \Sz \ .
\end{equation}
Next, we apply the $\Ucp$ rotation.  Here we use a Carr-Purcell (CP) series of $N$ equidistant $\pi$ pulses that is sandwiched between two $\pi/2$ rotations.  If the delay time $\tau$ between pulses is adjusted to one-half the precession period of the nuclear spin, $\tau = \pi/(\won+\apar/2)$, the $\pi$ pulse series generates a nuclear Rabi rotation with the effective Hamiltonian
\begin{equation}
H_N = \frac{\aperp}{\pi} \Sz\Ix 
\label{eq:HN}
\end{equation}
and simultaneously rotates the electronic spin around $z$ \cite{taminiau14}.  The overall effect of the $\Ucp$ sequence (including the $\pi/2$ pulses) is
\begin{equation}
\sigma_a = \sa\isa = \Ucp\sigma_0\iUcp = \mr{cos}\phi \Sx + \mr{sin}\phi \Sz\Ix \ ,
\label{eq:sigmaaa} 
\end{equation}
where $\phi = \aperp t/\pi$ and $t=N\tau$.  We notice that the $\Ucp$ rotation creates two components: An electronic coherence $\Sx$, and a nuclear coherence $\Sz\Ix$ that is conditional on the state of the electronic spin.  

We now consider the effect of the free nuclear precession.  The evolution $\Ufree=e^{-i\Hfree t_1}$ modulates both $\Sx$ and $\Sz\Ix$ as a function of time $t_1$, leading to oscillations in the signal $p$.  If we restrict ourselves to Hamiltonians $\Hfree$ that commute with $\Sz$ (see Fig. \ref{fig:fig1}c for examples), we can observe the separate evolutions of the electronic and nuclear spins, 
\begin{eqnarray}
p & = & |\isa \Ufree \sa|^2 = \tr( \Ufree \sigma_a \iUfree \sigma_a)  \\
	& = & \mr{cos}^2\phi \ \tr(\Ufree\Sx\iUfree\Sx) \nonumber \\
	&   & +\ \mr{sin}^2\phi \ \tr(\Ufree\Ix\iUfree\Ix) \ .
\end{eqnarray}
To detect the free precession signal, we therefore simply need to record $p$ as a function of time $t_1$ and pick out the desired nuclear ($\mr{sin}^2\phi$) part of the signal.  The latter can be achieved in several ways, for example by phase cycling (see Fig. \ref{fig:fig1}b) or by exploiting that the nuclear coherence is much longer-lived than the electronic coherence \cite{laraoui13}.

\begin{figure*}
\includegraphics[width=0.95\textwidth]{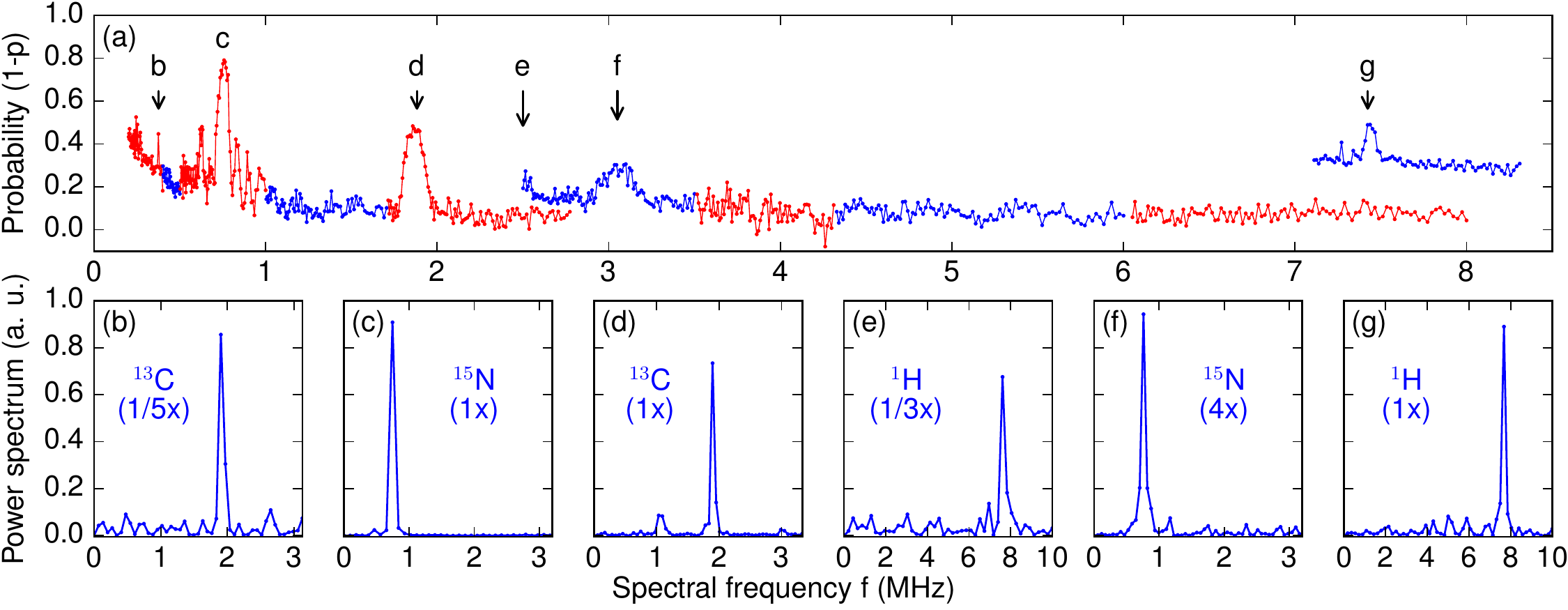}
\caption{\captionstyle
Unambiguous identification of NMR signals:
(a) Multipulse spectrum in a field of $B_0=174\unit{mT}$.  Colors identify different datasets.
(b-g) Fourier spectra for the six peaks labeled in (a).  All spectra used Hamiltonian $\Hfree^{(1)}$, except for (d,e) that used $\Hfree^{(2)}$ (see Fig. \ref{fig:fig1}c).  The nuclear species and multipulse harmonic are indicated for each spectrum.
}
\label{fig:fig2}
\end{figure*}
Experiments were performed on an electronic-grade diamond chip with a $\sim 5\unit{nm}$ deep layer of NV centers created by $^{15}$N$^+$ ion implantation at 2.5 keV (see Fig. \ref{fig:fig1}d).  The chip surface was patterned with an array of nanopillar waveguides \cite{babinec10} in order to increase the photon count rate.  The chip was integrated in a confocal fluorescence microscope equipped with a $532\unit{nm}$ excitation laser and a photon-counting module.  Laser pulses of $1.5\unit{\us}$ were used to initialize and readout the state of single NV electronic spins, and microwave pulses passed through a lithographically patterned transmission line were used for spin manipulation.  All measurements exploited the $m_S=0\leftrightarrow m_S=-1$ spin transition and were carried out in a bias field of approximately $0.18\unit{T}$ and at room temperature.
%The nuclear isotopes addressed in this study were \H, \C and \NN; their expected NMR frequencies are collected in Table \ref{table:species}.

As a first application, we show that the $\iUcp\Ufree\Ucp$ sequence can uniquely identify an NMR resonance and thereby resolve ambiguities with harmonics affecting traditional multipulse spectroscopy \cite{loretz15}.  To demonstrate this, we first record an extended multipulse spectrum using a simple $\Ucp$ sequence (Fig. \ref{fig:fig2}a).  The spectrum shows many peaks caused by the resonances of nuclear spins in the vicinity of the NV center.  Some of the peaks (c,d,g) can be tentatively assigned to \H, \C and \NN by consulting Table \ref{table:species}.  The other peaks (b,e,f) have frequencies that are not found in the Table, and it is not obvious which nuclear species they belong to.

Fig. \ref{fig:fig2}(b-g) shows spectra obtained with the $\iUcp\Ufree\Ucp$ sequence.  These spectra were recorded by tuning $\Ucp$ (via adjusting $\tau$) to one of the peaks labeled in (a), measuring the free precession for a series of $t_1$ values, and performing a Fourier transform with respect to $t_1$.  We find that all Fourier spectra show peaks at expected NMR frequencies (Table \ref{table:species}) and that we are now able to uniquely identify all nuclear species.

\begin{table}[b!]
\centering
\begin{tabular}{cccccc}
\hline\hline
\quad Isotope \quad & \quad NMR frequency (at 174 mT) \quad \\
\hline
\H      & $7.40\unit{MHz}$  \\
\C      & $1.86\unit{MHz}$  \\
\NN     & $0.75\unit{MHz}$, $2.30\unit{MHz}$ \\
\hline\hline
\end{tabular}
\caption{NMR frequencies for the nuclear spin species detected in the experiment. For $^{15}$N, two resonances are possible that are split by a hyperfine coupling of $3.05\unit{MHz}$.}
\label{table:species}
\end{table}
By comparing the frequencies between corresponding peaks in the multipulse and Fourier spectra we can discriminate different harmonics of the multipulse sequence.  We find that both ``ordinary'' harmonics created at $f/k$ and ``spurious'' harmonics appearing at $2f/k$ and $4f/k$ are present (where $k=1,3,5,...$).  Ordinary harmonics are expected resonances of the multipulse filter function \cite{cywinski08,delange11,zhao11} that lead to the same type of entanglement [Eq. (\ref{eq:sigmaaa})] as the fundamental signal.  Spurious harmonics, by contrast, result from phase accumulation during $\pi$ pulses and depend on many parameters, including $\pi$ pulse duration and amplitude, detuning, and the number of pulses \cite{loretz15}.  Despite the complicated evolution under $\Ucp$, spurious harmonics evidently generate some entanglement that leads to the appearance of a free precession signal.

\begin{figure}
\includegraphics[width=0.9\columnwidth]{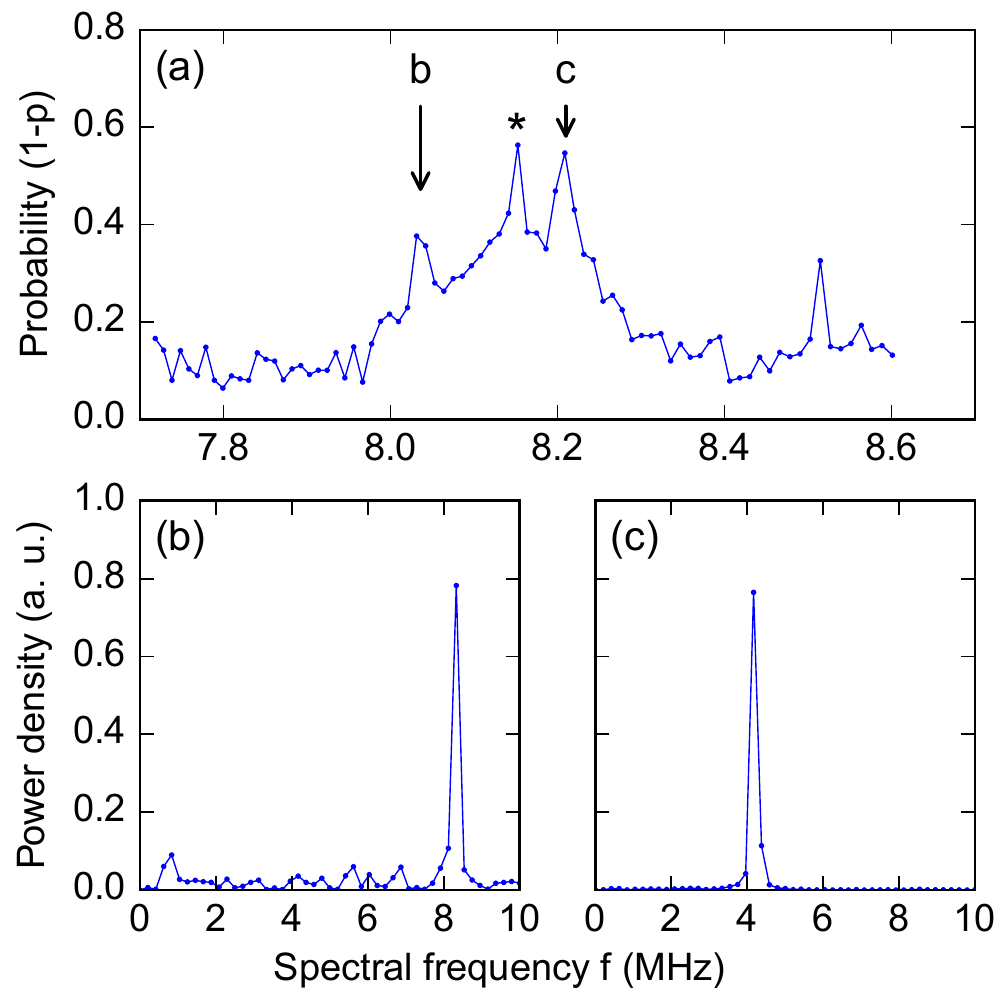}
\caption{\captionstyle
(a) Multipulse spectrum and (b,c) Fourier spectra at $\sim 8\unit{MHz}$.  Two signals are present, one produced by the fundamental resonance of a \H ensemble (b) and one produced by the second harmonic of a strongly-coupled \C nucleus (c).
The third peak ($\ast$) in (a) is caused by the same \C (data not shown). 
The number of pulses applied during $\Ucp$ was $N=$1200 in (a), 720 in (b) and 1000 in (c).
}
\label{fig:fig3}
\end{figure}
Fig. \ref{fig:fig3} shows another example of a multipulse spectrum alongside with two Fourier spectra.  In the multipulse spectrum, three peaks are visible that partially overlap.  The Fourier spectra reveal that the peaks are caused by the resonances of two nuclear species, including the second harmonic of a strongly-coupled \C and the fundamental signal of a proton ensemble.

As a second application, we demonstrate that the $\iUcp\Ufree\Ucp$ sequence allows for precise measurements of the nuclear Zeeman frequency $\won$ as well as the two hyperfine parameters $\apar$ and $\aperp$.  A precise measurement of $\won$ is important to assign the nuclear species and to resolve weak additional effects, such as chemical shifts or nuclear dipole couplings.  Accurate knowledge of $\apar$ and $\aperp$ is critical for determining the relative locations of nuclei in a molecule \cite{loretz14science}. 

\begin{figure}
\includegraphics[width=\columnwidth]{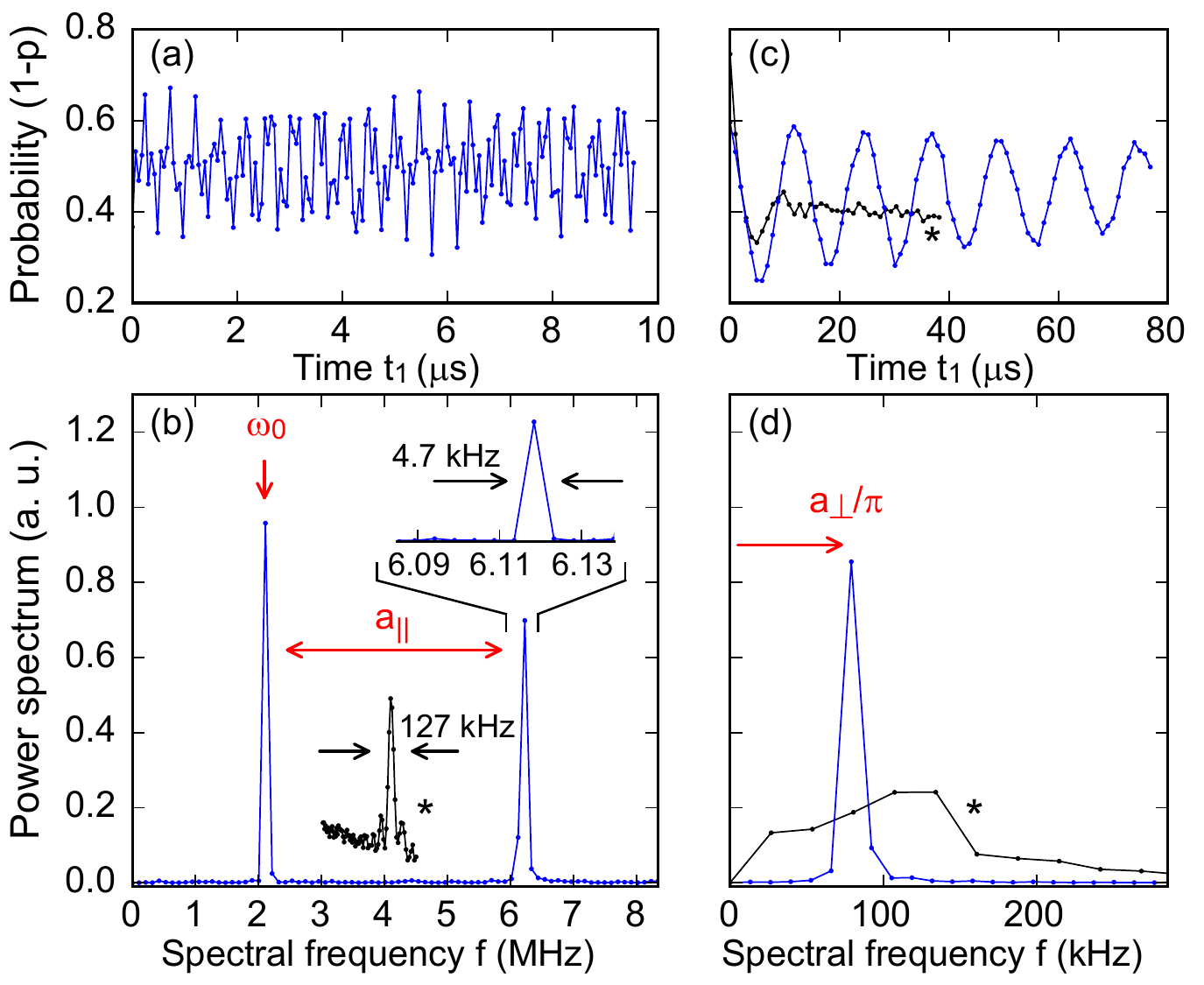}
\caption{\captionstyle
Precision measurement of $\won$, $\apar$ and $\aperp$ obtained by observing the nuclear, rather than the electronic, evolution.  The same \C as in Fig. \ref{fig:fig3} is being studied.
(a) Free precession signal under Hamiltonian $\Hfree^{(2)}$ (see Fig. \ref{fig:fig1}c).
(b) Fourier spectrum of (a) showing two peaks at $\won$ and $\won+\apar$.  The deduced parameters are $\won/2\pi = 2.09321(11)\unit{MHz}$ and $\apar/2\pi = 4.02350(16)\unit{MHz}$.  ($\ast$) shows a traditional multipulse spectrum for comparison.
(c) Free precession signal under Hamiltonian $\Hfree^{(3)}$ (see Fig. \ref{fig:fig1}c). ($\ast$) shows the corresponding multipulse signal obtained by incrementing the number of $\pi$ pulses $N=t_1/\tau$.
(d) Fourier spectra of the two curves from (c).  The deduced parameter is $\aperp/2\pi = 251.35(63)\unit{kHz}$.
}
\label{fig:fig4}
\end{figure}
To measure $\won$ and $\apar$, we perform a free precession experiment without the central $\pi$ pulse (Hamiltonian $\Hfree^{(2)}$ in Fig. \ref{fig:fig1}c).  The evolution $\Ufree$ now becomes conditional on the state of the electronic spin.  If the electron is in state $\ma$, the hyperfine interaction is absent and the nucleus precesses with frequency $\won$.  By contrast, if the electron is in state $\mb$, the hyperfine interaction is present and the nuclear precession frequency is $\won+\apar$.  Two peaks therefore appear reflecting these two frequencies (see Fig. \ref{fig:fig4}a,b).
From the peak positions one can directly deduce separate values for $\won$ and $\apar$.
By comparison, the conventional multipulse spectrum only shows a single, broad peak at $\won+\apar/2$ (see Figure).

Since the nuclear coherence is maintained for much longer than the electronic $T_2$ time (here $T_2\sim 10\unit{\us}$), the accuracy of the hyperfine measurement is greatly enhanced.  
The measurement precision is only limited by how long the free precession can be recorded, which in our experiments is set by the electronic $T_1$ time (here $T_1\sim 0.2\unit{ms}$) .
We have used undersampling to acquire nuclear precession signals for periods up to $t_1 = 200\unit{\us}$ (data not shown).  This has allowed us to determine $\won$ and $\apar$ with five digits of precision (see Figure).

By applying a train of $\pi$ pulses during the free precession period (Hamiltonian $\Hfree^{(3)}$) we can also measure the transverse coupling parameter $\aperp$ (see Fig. \ref{fig:fig4}c,d).  The $\pi$ pulses periodically flip the electronic spin between $\ma$ and $\mb$ and create a time-dependent hyperfine field that drives a nuclear Rabi oscillation with frequency $\aperp/\pi$, according to Eq. (\ref{eq:HN}).  Note that because the electronic spin is always in an eigenstate, no decoherence occurs and the Rabi oscillations can be driven up to the electronic $T_1$ time.  By contrast, the multipulse signal shows a rapid decay limited by the electronic coherence time $T_2$.

\begin{figure}
\includegraphics[width=\columnwidth]{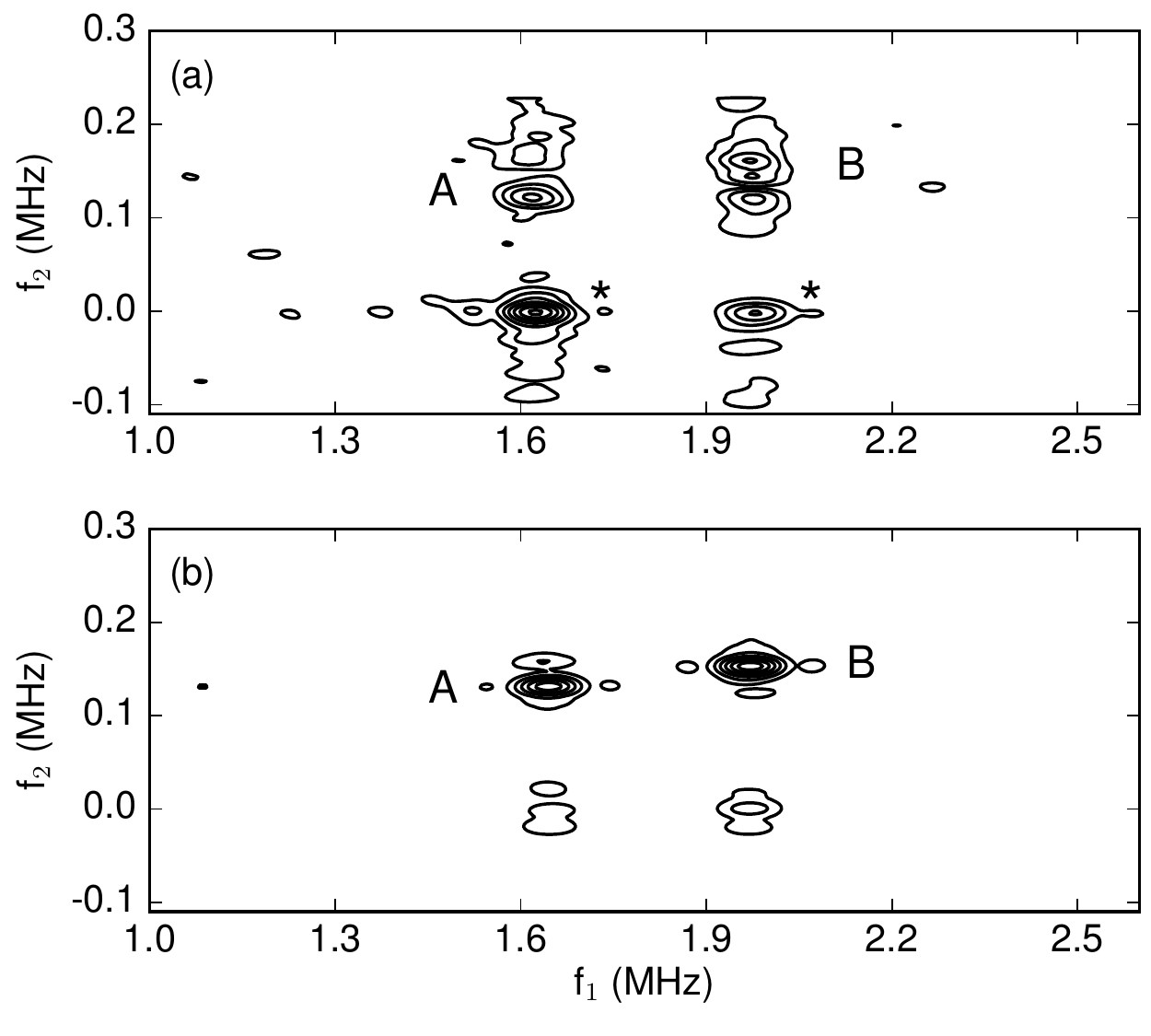}
\caption{\captionstyle
Two-dimensional Fourier NMR spectrum (absolute value) of two \C nuclei labeled by ``A'' and ``B''.
(a) is the measurement and (b) is a simulation.
The $f_1$ dimension plots $\won+\apar/2$ and the $f_2$ dimension plots $\aperp/\pi$.
Dwell time was $120\unit{ns}$ (120 increments) for $f_1$ and $2168\unit{ns}$ (40 increments) for $f_2$.
The peaks marked by $\ast$ are present because the phase of the nuclear Rabi rotation was not perfectly adjusted.
}
\label{fig:fig5}
\end{figure}

The two experiments can be combined to record two-dimensional (2D) Fourier spectra \cite{aue76}.
2D spectroscopy is an important technique in magnetic resonance for determining the connectivity in a spectrum \cite{ernst90} and it is an indispensable tool in biomolecular structure determination \cite{wuthrich03jbiolnmr}. 
Here we demonstrate that $\apar$ and $\aperp$ spectra can be correlated.
During a first period $t_1$ we apply Hamiltonian $\Hfree^{(1)}$ followed by a period $t_2$ of periodic $\pi$ pulses with Hamiltonian $\Hfree^{(3)}$ (see Fig. \ref{fig:fig1}c).  We then increment the evolution times $t_1$ and $t_2$ to produce a two-dimensional dataset that results, after a two-dimensional Fourier transform, in the spectrum shown in Fig. \ref{fig:fig5}.
Although only two \C are mapped in this measurement, we expect that 2D spectroscopy will be very useful if larger molecules with many nuclei are to be analyzed.

In summary, the concept of ``free nuclear precession'' has been implemented to demonstrate several important new capabilities in diamond-based NMR spectroscopy.  Spectral peaks could be uniquely assigned to nuclear isotopes, thereby resolving critical ambiguities with multipulse quantum sensing sequences.  The method further permitted selective measurements of the nuclear Zeeman frequency $\won$ and the hyperfine parameters $\apar$, $\aperp$ with 5 digits precision.  The precision could be extended further by storing the state of the electronic sensor spin in a nuclear quantum memory during free precession, for example in the $I=1/2$ \NN nucleus of the NV center \cite{laraoui13,rosskopf16}.  Two-dimensional Fourier spectroscopy was implemented and proposed as a potentially powerful tool for mapping nuclear coordinates in molecules.

%%%%%%%%%%%%%% Acknowledgments

The authors thank Matthew Markham and Andy Edmonds (ElementSix) for the custom diamond chip.
This work was supported by Swiss NSF Project Grant $200021\_137520$, the NCCR QSIT, and the DIADEMS programme 611143 of the European Commission.

\noindent

%\bibliography{library}
%\bibliography{C:/Christian/ETH/labview/library/library}

\begin{thebibliography}{28}%
\makeatletter
\providecommand \@ifxundefined [1]{%
 \@ifx{#1\undefined}
}%
\providecommand \@ifnum [1]{%
 \ifnum #1\expandafter \@firstoftwo
 \else \expandafter \@secondoftwo
 \fi
}%
\providecommand \@ifx [1]{%
 \ifx #1\expandafter \@firstoftwo
 \else \expandafter \@secondoftwo
 \fi
}%
\providecommand \natexlab [1]{#1}%
\providecommand \enquote  [1]{``#1''}%
\providecommand \bibnamefont  [1]{#1}%
\providecommand \bibfnamefont [1]{#1}%
\providecommand \citenamefont [1]{#1}%
\providecommand \href@noop [0]{\@secondoftwo}%
\providecommand \href [0]{\begingroup \@sanitize@url \@href}%
\providecommand \@href[1]{\@@startlink{#1}\@@href}%
\providecommand \@@href[1]{\endgroup#1\@@endlink}%
\providecommand \@sanitize@url [0]{\catcode `\\12\catcode `\$12\catcode
  `\&12\catcode `\#12\catcode `\^12\catcode `\_12\catcode `\%12\relax}%
\providecommand \@@startlink[1]{}%
\providecommand \@@endlink[0]{}%
\providecommand \url  [0]{\begingroup\@sanitize@url \@url }%
\providecommand \@url [1]{\endgroup\@href {#1}{\urlprefix }}%
\providecommand \urlprefix  [0]{URL }%
\providecommand \Eprint [0]{\href }%
\providecommand \doibase [0]{http://dx.doi.org/}%
\providecommand \selectlanguage [0]{\@gobble}%
\providecommand \bibinfo  [0]{\@secondoftwo}%
\providecommand \bibfield  [0]{\@secondoftwo}%
\providecommand \translation [1]{[#1]}%
\providecommand \BibitemOpen [0]{}%
\providecommand \bibitemStop [0]{}%
\providecommand \bibitemNoStop [0]{.\EOS\space}%
\providecommand \EOS [0]{\spacefactor3000\relax}%
\providecommand \BibitemShut  [1]{\csname bibitem#1\endcsname}%
\let\auto@bib@innerbib\@empty
%</preamble>
\bibitem [{\citenamefont {Degen}(2008)}]{degen08apl}%
  \BibitemOpen
  \bibfield  {author} {\bibinfo {author} {\bibfnamefont {C.~L.}\ \bibnamefont
  {Degen}},\ }\href {\doibase 10.1063/1.2943282} {\bibfield  {journal}
  {\bibinfo  {journal} {Appl. Phys. Lett.}\ }\textbf {\bibinfo {volume} {92}},\
  \bibinfo {eid} {243111} (\bibinfo {year} {2008})}\BibitemShut {NoStop}%
\bibitem [{\citenamefont {Mamin}\ \emph {et~al.}(2013)\citenamefont {Mamin},
  \citenamefont {Kim}, \citenamefont {Sherwood}, \citenamefont {Rettner},
  \citenamefont {Ohno}, \citenamefont {Awschalom},\ and\ \citenamefont
  {Rugar}}]{mamin13}%
  \BibitemOpen
  \bibfield  {author} {\bibinfo {author} {\bibfnamefont {H.~J.}\ \bibnamefont
  {Mamin}}, \bibinfo {author} {\bibfnamefont {M.}~\bibnamefont {Kim}}, \bibinfo
  {author} {\bibfnamefont {M.~H.}\ \bibnamefont {Sherwood}}, \bibinfo {author}
  {\bibfnamefont {C.~T.}\ \bibnamefont {Rettner}}, \bibinfo {author}
  {\bibfnamefont {K.}~\bibnamefont {Ohno}}, \bibinfo {author} {\bibfnamefont
  {D.~D.}\ \bibnamefont {Awschalom}}, \ and\ \bibinfo {author} {\bibfnamefont
  {D.}~\bibnamefont {Rugar}},\ }\href {\doibase 10.1126/science.1231540}
  {\bibfield  {journal} {\bibinfo  {journal} {Science}\ }\textbf {\bibinfo
  {volume} {339}},\ \bibinfo {pages} {557} (\bibinfo {year}
  {2013})}\BibitemShut {NoStop}%
\bibitem [{\citenamefont {Staudacher}\ \emph {et~al.}(2013)\citenamefont
  {Staudacher}, \citenamefont {Shi}, \citenamefont {Pezzagna}, \citenamefont
  {Meijer}, \citenamefont {Du}, \citenamefont {Meriles}, \citenamefont
  {Reinhard},\ and\ \citenamefont {Wrachtrup}}]{staudacher13}%
  \BibitemOpen
  \bibfield  {author} {\bibinfo {author} {\bibfnamefont {T.}~\bibnamefont
  {Staudacher}}, \bibinfo {author} {\bibfnamefont {F.}~\bibnamefont {Shi}},
  \bibinfo {author} {\bibfnamefont {S.}~\bibnamefont {Pezzagna}}, \bibinfo
  {author} {\bibfnamefont {J.}~\bibnamefont {Meijer}}, \bibinfo {author}
  {\bibfnamefont {J.}~\bibnamefont {Du}}, \bibinfo {author} {\bibfnamefont
  {C.~A.}\ \bibnamefont {Meriles}}, \bibinfo {author} {\bibfnamefont
  {F.}~\bibnamefont {Reinhard}}, \ and\ \bibinfo {author} {\bibfnamefont
  {J.}~\bibnamefont {Wrachtrup}},\ }\href {\doibase 10.1126/science.1231675}
  {\bibfield  {journal} {\bibinfo  {journal} {Science}\ }\textbf {\bibinfo
  {volume} {339}},\ \bibinfo {pages} {561} (\bibinfo {year}
  {2013})}\BibitemShut {NoStop}%
\bibitem [{\citenamefont {Davies}\ and\ \citenamefont
  {Hamer}(1976)}]{davies76}%
  \BibitemOpen
  \bibfield  {author} {\bibinfo {author} {\bibfnamefont {G.}~\bibnamefont
  {Davies}}\ and\ \bibinfo {author} {\bibfnamefont {M.~F.}\ \bibnamefont
  {Hamer}},\ }\href {\doibase 10.1098/rspa.1976.0039} {\bibfield  {journal}
  {\bibinfo  {journal} {Proc. Royal Soc. London A}\ }\textbf {\bibinfo {volume}
  {348}},\ \bibinfo {pages} {285} (\bibinfo {year} {1976})}\BibitemShut
  {NoStop}%
\bibitem [{\citenamefont {Gruber}\ \emph {et~al.}(1997)\citenamefont {Gruber},
  \citenamefont {Drabenstedt}, \citenamefont {Tietz}, \citenamefont {Fleury},
  \citenamefont {Wrachtrup},\ and\ \citenamefont {von
  Borczyskowski}}]{gruber97}%
  \BibitemOpen
  \bibfield  {author} {\bibinfo {author} {\bibfnamefont {A.}~\bibnamefont
  {Gruber}}, \bibinfo {author} {\bibfnamefont {A.}~\bibnamefont {Drabenstedt}},
  \bibinfo {author} {\bibfnamefont {C.}~\bibnamefont {Tietz}}, \bibinfo
  {author} {\bibfnamefont {L.}~\bibnamefont {Fleury}}, \bibinfo {author}
  {\bibfnamefont {J.}~\bibnamefont {Wrachtrup}}, \ and\ \bibinfo {author}
  {\bibfnamefont {C.}~\bibnamefont {von Borczyskowski}},\ }\href {\doibase
  10.1126/science.276.5321.2012} {\bibfield  {journal} {\bibinfo  {journal}
  {Science}\ }\textbf {\bibinfo {volume} {276}},\ \bibinfo {eid} {2012}
  (\bibinfo {year} {1997})}\BibitemShut {NoStop}%
\bibitem [{\citenamefont {Loretz}\ \emph
  {et~al.}(2014{\natexlab{a}})\citenamefont {Loretz}, \citenamefont {Pezzagna},
  \citenamefont {Meijer},\ and\ \citenamefont {Degen}}]{loretz14apl}%
  \BibitemOpen
  \bibfield  {author} {\bibinfo {author} {\bibfnamefont {M.}~\bibnamefont
  {Loretz}}, \bibinfo {author} {\bibfnamefont {S.}~\bibnamefont {Pezzagna}},
  \bibinfo {author} {\bibfnamefont {J.}~\bibnamefont {Meijer}}, \ and\ \bibinfo
  {author} {\bibfnamefont {C.~L.}\ \bibnamefont {Degen}},\ }\href {\doibase
  10.1063/1.4862749} {\bibfield  {journal} {\bibinfo  {journal} {Appl. Phys.
  Lett.}\ }\textbf {\bibinfo {volume} {104}},\ \bibinfo {pages} {33102}
  (\bibinfo {year} {2014}{\natexlab{a}})}\BibitemShut {NoStop}%
\bibitem [{\citenamefont {Muller}\ \emph {et~al.}(2014)\citenamefont {Muller},
  \citenamefont {Kong}, \citenamefont {Cai}, \citenamefont {Melentijevic},
  \citenamefont {Stacey}, \citenamefont {Markham}, \citenamefont {Twitchen},
  \citenamefont {Isoya}, \citenamefont {Pezzagna}, \citenamefont {Meijer},
  \citenamefont {Du}, \citenamefont {Plenio}, \citenamefont {Naydenov},
  \citenamefont {McGuinness},\ and\ \citenamefont {Jelezko}}]{muller14}%
  \BibitemOpen
  \bibfield  {author} {\bibinfo {author} {\bibfnamefont {C.}~\bibnamefont
  {Muller}}, \bibinfo {author} {\bibfnamefont {X.}~\bibnamefont {Kong}},
  \bibinfo {author} {\bibfnamefont {J.~M.}\ \bibnamefont {Cai}}, \bibinfo
  {author} {\bibfnamefont {K.}~\bibnamefont {Melentijevic}}, \bibinfo {author}
  {\bibfnamefont {A.}~\bibnamefont {Stacey}}, \bibinfo {author} {\bibfnamefont
  {M.}~\bibnamefont {Markham}}, \bibinfo {author} {\bibfnamefont
  {D.}~\bibnamefont {Twitchen}}, \bibinfo {author} {\bibfnamefont
  {J.}~\bibnamefont {Isoya}}, \bibinfo {author} {\bibfnamefont
  {S.}~\bibnamefont {Pezzagna}}, \bibinfo {author} {\bibfnamefont
  {J.}~\bibnamefont {Meijer}}, \bibinfo {author} {\bibfnamefont {J.~F.}\
  \bibnamefont {Du}}, \bibinfo {author} {\bibfnamefont {M.~B.}\ \bibnamefont
  {Plenio}}, \bibinfo {author} {\bibfnamefont {B.}~\bibnamefont {Naydenov}},
  \bibinfo {author} {\bibfnamefont {L.~P.}\ \bibnamefont {McGuinness}}, \ and\
  \bibinfo {author} {\bibfnamefont {F.}~\bibnamefont {Jelezko}},\ }\href
  {\doibase 10.1038/ncomms5703} {\bibfield  {journal} {\bibinfo  {journal}
  {Nature Commun.}\ }\textbf {\bibinfo {volume} {5}},\ \bibinfo {pages} {4703}
  (\bibinfo {year} {2014})}\BibitemShut {NoStop}%
\bibitem [{\citenamefont {Devience}\ \emph {et~al.}(2015)\citenamefont
  {Devience}, \citenamefont {Pham}, \citenamefont {Lovchinsky}, \citenamefont
  {Sushkov}, \citenamefont {Bar-gill}, \citenamefont {Belthangady},
  \citenamefont {Casola}, \citenamefont {Corbett}, \citenamefont {Zhang},
  \citenamefont {Lukin}, \citenamefont {Park}, \citenamefont {Yacoby},\ and\
  \citenamefont {Walsworth}}]{devience15}%
  \BibitemOpen
  \bibfield  {author} {\bibinfo {author} {\bibfnamefont {S.~J.}\ \bibnamefont
  {Devience}}, \bibinfo {author} {\bibfnamefont {L.~M.}\ \bibnamefont {Pham}},
  \bibinfo {author} {\bibfnamefont {I.}~\bibnamefont {Lovchinsky}}, \bibinfo
  {author} {\bibfnamefont {A.~O.}\ \bibnamefont {Sushkov}}, \bibinfo {author}
  {\bibfnamefont {N.}~\bibnamefont {Bar-gill}}, \bibinfo {author}
  {\bibfnamefont {C.}~\bibnamefont {Belthangady}}, \bibinfo {author}
  {\bibfnamefont {F.}~\bibnamefont {Casola}}, \bibinfo {author} {\bibfnamefont
  {M.}~\bibnamefont {Corbett}}, \bibinfo {author} {\bibfnamefont
  {H.}~\bibnamefont {Zhang}}, \bibinfo {author} {\bibfnamefont
  {M.}~\bibnamefont {Lukin}}, \bibinfo {author} {\bibfnamefont
  {H.}~\bibnamefont {Park}}, \bibinfo {author} {\bibfnamefont {A.}~\bibnamefont
  {Yacoby}}, \ and\ \bibinfo {author} {\bibfnamefont {R.~L.}\ \bibnamefont
  {Walsworth}},\ }\href {\doibase 10.1038/nnano.2014.313} {\bibfield  {journal}
  {\bibinfo  {journal} {Nature Nano.}\ }\textbf {\bibinfo {volume} {10}},\
  \bibinfo {pages} {129} (\bibinfo {year} {2015})}\BibitemShut {NoStop}%
\bibitem [{\citenamefont {Haberle}\ \emph {et~al.}(2015)\citenamefont
  {Haberle}, \citenamefont {Schmid-Lorch}, \citenamefont {Reinhard},\ and\
  \citenamefont {Wrachtrup}}]{haberle15}%
  \BibitemOpen
  \bibfield  {author} {\bibinfo {author} {\bibfnamefont {T.}~\bibnamefont
  {Haberle}}, \bibinfo {author} {\bibfnamefont {D.}~\bibnamefont
  {Schmid-Lorch}}, \bibinfo {author} {\bibfnamefont {F.}~\bibnamefont
  {Reinhard}}, \ and\ \bibinfo {author} {\bibfnamefont {J.}~\bibnamefont
  {Wrachtrup}},\ }\href {\doibase 10.1038/nnano.2014.299} {\bibfield  {journal}
  {\bibinfo  {journal} {Nature Nano.}\ }\textbf {\bibinfo {volume} {10}},\
  \bibinfo {pages} {125} (\bibinfo {year} {2015})}\BibitemShut {NoStop}%
\bibitem [{\citenamefont {Kong}\ \emph {et~al.}(2015)\citenamefont {Kong},
  \citenamefont {Stark}, \citenamefont {Du}, \citenamefont {McGuinness},\ and\
  \citenamefont {Jelezko}}]{kong15}%
  \BibitemOpen
  \bibfield  {author} {\bibinfo {author} {\bibfnamefont {X.}~\bibnamefont
  {Kong}}, \bibinfo {author} {\bibfnamefont {A.}~\bibnamefont {Stark}},
  \bibinfo {author} {\bibfnamefont {J.}~\bibnamefont {Du}}, \bibinfo {author}
  {\bibfnamefont {L.~P.}\ \bibnamefont {McGuinness}}, \ and\ \bibinfo {author}
  {\bibfnamefont {F.}~\bibnamefont {Jelezko}},\ }\href {\doibase
  10.1103/PhysRevApplied.4.024004} {\bibfield  {journal} {\bibinfo  {journal}
  {Phys. Rev. Applied}\ }\textbf {\bibinfo {volume} {4}},\ \bibinfo {pages}
  {024004} (\bibinfo {year} {2015})}\BibitemShut {NoStop}%
\bibitem [{\citenamefont {Staudacher}\ \emph {et~al.}(2015)\citenamefont
  {Staudacher}, \citenamefont {Raatz}, \citenamefont {Pezzagna}, \citenamefont
  {Meijer}, \citenamefont {Reinhard}, \citenamefont {Meriles},\ and\
  \citenamefont {Wrachtrup}}]{staudacher15}%
  \BibitemOpen
  \bibfield  {author} {\bibinfo {author} {\bibfnamefont {T.}~\bibnamefont
  {Staudacher}}, \bibinfo {author} {\bibfnamefont {N.}~\bibnamefont {Raatz}},
  \bibinfo {author} {\bibfnamefont {S.}~\bibnamefont {Pezzagna}}, \bibinfo
  {author} {\bibfnamefont {J.}~\bibnamefont {Meijer}}, \bibinfo {author}
  {\bibfnamefont {F.}~\bibnamefont {Reinhard}}, \bibinfo {author}
  {\bibfnamefont {C.~A.}\ \bibnamefont {Meriles}}, \ and\ \bibinfo {author}
  {\bibfnamefont {J.}~\bibnamefont {Wrachtrup}},\ }\href {\doibase
  10.1038/ncomms9527} {\bibfield  {journal} {\bibinfo  {journal} {Nature
  Commun.}\ }\textbf {\bibinfo {volume} {6}} (\bibinfo {year} {2015}),\
  10.1038/ncomms9527}\BibitemShut {NoStop}%
\bibitem [{\citenamefont {Rugar}\ \emph {et~al.}(2015)\citenamefont {Rugar},
  \citenamefont {Mamin}, \citenamefont {Sherwood}, \citenamefont {Kim},
  \citenamefont {Rettner}, \citenamefont {Ohno},\ and\ \citenamefont
  {Awschalom}}]{rugar15}%
  \BibitemOpen
  \bibfield  {author} {\bibinfo {author} {\bibfnamefont {D.}~\bibnamefont
  {Rugar}}, \bibinfo {author} {\bibfnamefont {H.~J.}\ \bibnamefont {Mamin}},
  \bibinfo {author} {\bibfnamefont {M.~H.}\ \bibnamefont {Sherwood}}, \bibinfo
  {author} {\bibfnamefont {M.}~\bibnamefont {Kim}}, \bibinfo {author}
  {\bibfnamefont {C.~T.}\ \bibnamefont {Rettner}}, \bibinfo {author}
  {\bibfnamefont {K.}~\bibnamefont {Ohno}}, \ and\ \bibinfo {author}
  {\bibfnamefont {D.~D.}\ \bibnamefont {Awschalom}},\ }\href {\doibase
  10.1038/nnano.2014.288} {\bibfield  {journal} {\bibinfo  {journal} {Nature
  Nano.}\ }\textbf {\bibinfo {volume} {10}},\ \bibinfo {pages} {120} (\bibinfo
  {year} {2015})}\BibitemShut {NoStop}%
\bibitem [{\citenamefont {Schirhagl}\ \emph {et~al.}(2014)\citenamefont
  {Schirhagl}, \citenamefont {Chang}, \citenamefont {Loretz},\ and\
  \citenamefont {Degen}}]{schirhagl14}%
  \BibitemOpen
  \bibfield  {author} {\bibinfo {author} {\bibfnamefont {R.}~\bibnamefont
  {Schirhagl}}, \bibinfo {author} {\bibfnamefont {K.}~\bibnamefont {Chang}},
  \bibinfo {author} {\bibfnamefont {M.}~\bibnamefont {Loretz}}, \ and\ \bibinfo
  {author} {\bibfnamefont {C.~L.}\ \bibnamefont {Degen}},\ }\href {\doibase
  10.1146/annurev-physchem-040513-103659} {\bibfield  {journal} {\bibinfo
  {journal} {Annu. Rev. Phys. Chem.}\ }\textbf {\bibinfo {volume} {65}},\
  \bibinfo {pages} {83} (\bibinfo {year} {2014})}\BibitemShut {NoStop}%
\bibitem [{\citenamefont {Zhao}\ \emph {et~al.}(2011)\citenamefont {Zhao},
  \citenamefont {Hu}, \citenamefont {Ho}, \citenamefont {Wan},\ and\
  \citenamefont {Liu}}]{zhao11}%
  \BibitemOpen
  \bibfield  {author} {\bibinfo {author} {\bibfnamefont {N.}~\bibnamefont
  {Zhao}}, \bibinfo {author} {\bibfnamefont {J.~L.}\ \bibnamefont {Hu}},
  \bibinfo {author} {\bibfnamefont {S.~W.}\ \bibnamefont {Ho}}, \bibinfo
  {author} {\bibfnamefont {J.~T.~K.}\ \bibnamefont {Wan}}, \ and\ \bibinfo
  {author} {\bibfnamefont {R.~B.}\ \bibnamefont {Liu}},\ }\href {\doibase
  10.1038/nnano.2011.22} {\bibfield  {journal} {\bibinfo  {journal} {Nature
  Nanotechnology}\ }\textbf {\bibinfo {volume} {6}},\ \bibinfo {pages} {242}
  (\bibinfo {year} {2011})}\BibitemShut {NoStop}%
\bibitem [{\citenamefont {Kotler}\ \emph {et~al.}(2011)\citenamefont {Kotler},
  \citenamefont {Akerman}, \citenamefont {Glickman}, \citenamefont {Keselman},\
  and\ \citenamefont {Ozeri}}]{kotler11}%
  \BibitemOpen
  \bibfield  {author} {\bibinfo {author} {\bibfnamefont {S.}~\bibnamefont
  {Kotler}}, \bibinfo {author} {\bibfnamefont {N.}~\bibnamefont {Akerman}},
  \bibinfo {author} {\bibfnamefont {Y.}~\bibnamefont {Glickman}}, \bibinfo
  {author} {\bibfnamefont {A.}~\bibnamefont {Keselman}}, \ and\ \bibinfo
  {author} {\bibfnamefont {R.}~\bibnamefont {Ozeri}},\ }\href {\doibase
  10.1038/nature10010} {\bibfield  {journal} {\bibinfo  {journal} {Nature}\
  }\textbf {\bibinfo {volume} {473}},\ \bibinfo {pages} {61} (\bibinfo {year}
  {2011})}\BibitemShut {NoStop}%
\bibitem [{\citenamefont {Cywinski}\ \emph {et~al.}(2008)\citenamefont
  {Cywinski}, \citenamefont {Lutchyn}, \citenamefont {Nave},\ and\
  \citenamefont {Sarma}}]{cywinski08}%
  \BibitemOpen
  \bibfield  {author} {\bibinfo {author} {\bibfnamefont {L.}~\bibnamefont
  {Cywinski}}, \bibinfo {author} {\bibfnamefont {R.~M.}\ \bibnamefont
  {Lutchyn}}, \bibinfo {author} {\bibfnamefont {C.~P.}\ \bibnamefont {Nave}}, \
  and\ \bibinfo {author} {\bibfnamefont {S.~D.}\ \bibnamefont {Sarma}},\ }\href
  {\doibase 10.1103/PhysRevB.77.174509} {\bibfield  {journal} {\bibinfo
  {journal} {Phys. Rev. B}\ }\textbf {\bibinfo {volume} {77}},\ \bibinfo
  {pages} {174509} (\bibinfo {year} {2008})}\BibitemShut {NoStop}%
\bibitem [{\citenamefont {Lange}\ \emph {et~al.}(2011)\citenamefont {Lange},
  \citenamefont {Riste}, \citenamefont {Dobrovitski},\ and\ \citenamefont
  {Hanson}}]{delange11}%
  \BibitemOpen
  \bibfield  {author} {\bibinfo {author} {\bibfnamefont {G.~D.}\ \bibnamefont
  {Lange}}, \bibinfo {author} {\bibfnamefont {D.}~\bibnamefont {Riste}},
  \bibinfo {author} {\bibfnamefont {V.~V.}\ \bibnamefont {Dobrovitski}}, \ and\
  \bibinfo {author} {\bibfnamefont {R.}~\bibnamefont {Hanson}},\ }\href
  {\doibase 10.1103/PhysRevLett.106.080802} {\bibfield  {journal} {\bibinfo
  {journal} {Phys. Rev. Lett.}\ }\textbf {\bibinfo {volume} {106}},\ \bibinfo
  {pages} {080802} (\bibinfo {year} {2011})}\BibitemShut {NoStop}%
\bibitem [{\citenamefont {Loretz}\ \emph {et~al.}(2015)\citenamefont {Loretz},
  \citenamefont {Boss}, \citenamefont {Rosskopf}, \citenamefont {Mamin},
  \citenamefont {Rugar},\ and\ \citenamefont {Degen}}]{loretz15}%
  \BibitemOpen
  \bibfield  {author} {\bibinfo {author} {\bibfnamefont {M.}~\bibnamefont
  {Loretz}}, \bibinfo {author} {\bibfnamefont {J.~M.}\ \bibnamefont {Boss}},
  \bibinfo {author} {\bibfnamefont {T.}~\bibnamefont {Rosskopf}}, \bibinfo
  {author} {\bibfnamefont {H.~J.}\ \bibnamefont {Mamin}}, \bibinfo {author}
  {\bibfnamefont {D.}~\bibnamefont {Rugar}}, \ and\ \bibinfo {author}
  {\bibfnamefont {C.~L.}\ \bibnamefont {Degen}},\ }\href {\doibase
  10.1103/PhysRevX.5.021009} {\bibfield  {journal} {\bibinfo  {journal} {Phys.
  Rev. X}\ }\textbf {\bibinfo {volume} {5}},\ \bibinfo {pages} {21009}
  (\bibinfo {year} {2015})}\BibitemShut {NoStop}%
\bibitem [{\citenamefont {Laraoui}\ \emph {et~al.}(2013)\citenamefont
  {Laraoui}, \citenamefont {Dolde}, \citenamefont {Burk}, \citenamefont
  {Reinhard}, \citenamefont {Wrachtrup},\ and\ \citenamefont
  {Meriles}}]{laraoui13}%
  \BibitemOpen
  \bibfield  {author} {\bibinfo {author} {\bibfnamefont {A.}~\bibnamefont
  {Laraoui}}, \bibinfo {author} {\bibfnamefont {F.}~\bibnamefont {Dolde}},
  \bibinfo {author} {\bibfnamefont {C.}~\bibnamefont {Burk}}, \bibinfo {author}
  {\bibfnamefont {F.}~\bibnamefont {Reinhard}}, \bibinfo {author}
  {\bibfnamefont {J.}~\bibnamefont {Wrachtrup}}, \ and\ \bibinfo {author}
  {\bibfnamefont {C.~A.}\ \bibnamefont {Meriles}},\ }\href {\doibase
  10.1038/ncomms2685} {\bibfield  {journal} {\bibinfo  {journal} {Nature
  Commun.}\ }\textbf {\bibinfo {volume} {4}},\ \bibinfo {pages} {1651}
  (\bibinfo {year} {2013})}\BibitemShut {NoStop}%
\bibitem [{\citenamefont {Gullion}\ \emph {et~al.}(1990)\citenamefont
  {Gullion}, \citenamefont {Baker},\ and\ \citenamefont {Conradi}}]{gullion90}%
  \BibitemOpen
  \bibfield  {author} {\bibinfo {author} {\bibfnamefont {T.}~\bibnamefont
  {Gullion}}, \bibinfo {author} {\bibfnamefont {D.~B.}\ \bibnamefont {Baker}},
  \ and\ \bibinfo {author} {\bibfnamefont {M.~S.}\ \bibnamefont {Conradi}},\
  }\href {http://www.sciencedirect.com/science/article/pii/0022236490903313}
  {\bibfield  {journal} {\bibinfo  {journal} {J. Magn. Res.}\ }\textbf
  {\bibinfo {volume} {89}},\ \bibinfo {pages} {479} (\bibinfo {year}
  {1990})}\BibitemShut {NoStop}%
\bibitem [{\citenamefont {Taminiau}\ \emph {et~al.}(2012)\citenamefont
  {Taminiau}, \citenamefont {Wagenaar}, \citenamefont {der Sar}, \citenamefont
  {Jelezko}, \citenamefont {Dobrovitski},\ and\ \citenamefont
  {Hanson}}]{taminiau12}%
  \BibitemOpen
  \bibfield  {author} {\bibinfo {author} {\bibfnamefont {T.~H.}\ \bibnamefont
  {Taminiau}}, \bibinfo {author} {\bibfnamefont {J.~J.~T.}\ \bibnamefont
  {Wagenaar}}, \bibinfo {author} {\bibfnamefont {T.~V.}\ \bibnamefont {der
  Sar}}, \bibinfo {author} {\bibfnamefont {F.}~\bibnamefont {Jelezko}},
  \bibinfo {author} {\bibfnamefont {V.~V.}\ \bibnamefont {Dobrovitski}}, \ and\
  \bibinfo {author} {\bibfnamefont {R.}~\bibnamefont {Hanson}},\ }\href
  {\doibase 10.1103/PhysRevLett.109.137602} {\bibfield  {journal} {\bibinfo
  {journal} {Phys. Rev. Lett.}\ }\textbf {\bibinfo {volume} {109}},\ \bibinfo
  {pages} {137602} (\bibinfo {year} {2012})}\BibitemShut {NoStop}%
\bibitem [{\citenamefont {Loretz}\ \emph
  {et~al.}(2014{\natexlab{b}})\citenamefont {Loretz}, \citenamefont {Rosskopf},
  \citenamefont {Boss}, \citenamefont {Pezzagna}, \citenamefont {Meijer},\ and\
  \citenamefont {Degen}}]{loretz14science}%
  \BibitemOpen
  \bibfield  {author} {\bibinfo {author} {\bibfnamefont {M.}~\bibnamefont
  {Loretz}}, \bibinfo {author} {\bibfnamefont {T.}~\bibnamefont {Rosskopf}},
  \bibinfo {author} {\bibfnamefont {J.~M.}\ \bibnamefont {Boss}}, \bibinfo
  {author} {\bibfnamefont {S.}~\bibnamefont {Pezzagna}}, \bibinfo {author}
  {\bibfnamefont {J.}~\bibnamefont {Meijer}}, \ and\ \bibinfo {author}
  {\bibfnamefont {C.~L.}\ \bibnamefont {Degen}},\ }\href {\doibase
  10.1126/science.1259464} {\bibfield  {journal} {\bibinfo  {journal}
  {Science}\ } (\bibinfo {year} {2014}{\natexlab{b}}),\
  10.1126/science.1259464}\BibitemShut {NoStop}%
\bibitem [{\citenamefont {Aue}\ \emph {et~al.}(1976)\citenamefont {Aue},
  \citenamefont {Bartholdi},\ and\ \citenamefont {Ernst}}]{aue76}%
  \BibitemOpen
  \bibfield  {author} {\bibinfo {author} {\bibfnamefont {W.~P.}\ \bibnamefont
  {Aue}}, \bibinfo {author} {\bibfnamefont {E.}~\bibnamefont {Bartholdi}}, \
  and\ \bibinfo {author} {\bibfnamefont {R.~R.}\ \bibnamefont {Ernst}},\ }\href
  {\doibase 10.1063/1.432450} {\bibfield  {journal} {\bibinfo  {journal} {J.
  Chem. Phys.}\ }\textbf {\bibinfo {volume} {64}},\ \bibinfo {pages} {2229}
  (\bibinfo {year} {1976})}\BibitemShut {NoStop}%
\bibitem [{\citenamefont {Taminiau}\ \emph {et~al.}(2014)\citenamefont
  {Taminiau}, \citenamefont {Cramer}, \citenamefont {van~der Sar},
  \citenamefont {Dobrovitski},\ and\ \citenamefont {Hanson}}]{taminiau14}%
  \BibitemOpen
  \bibfield  {author} {\bibinfo {author} {\bibfnamefont {T.~H.}\ \bibnamefont
  {Taminiau}}, \bibinfo {author} {\bibfnamefont {J.}~\bibnamefont {Cramer}},
  \bibinfo {author} {\bibfnamefont {T.}~\bibnamefont {van~der Sar}}, \bibinfo
  {author} {\bibfnamefont {V.~V.}\ \bibnamefont {Dobrovitski}}, \ and\ \bibinfo
  {author} {\bibfnamefont {R.}~\bibnamefont {Hanson}},\ }\href {\doibase
  10.1038/nnano.2014.2} {\bibfield  {journal} {\bibinfo  {journal} {Nature
  Nano.}\ }\textbf {\bibinfo {volume} {9}} (\bibinfo {year} {2014}),\
  10.1038/nnano.2014.2}\BibitemShut {NoStop}%
\bibitem [{\citenamefont {Babinec}\ \emph {et~al.}(2010)\citenamefont
  {Babinec}, \citenamefont {Hausmann}, \citenamefont {Khan}, \citenamefont
  {Zhang}, \citenamefont {Maze}, \citenamefont {Hemmer},\ and\ \citenamefont
  {Loncar}}]{babinec10}%
  \BibitemOpen
  \bibfield  {author} {\bibinfo {author} {\bibfnamefont {T.~M.}\ \bibnamefont
  {Babinec}}, \bibinfo {author} {\bibfnamefont {B.~J.~M.}\ \bibnamefont
  {Hausmann}}, \bibinfo {author} {\bibfnamefont {M.}~\bibnamefont {Khan}},
  \bibinfo {author} {\bibfnamefont {Y.}~\bibnamefont {Zhang}}, \bibinfo
  {author} {\bibfnamefont {J.~R.}\ \bibnamefont {Maze}}, \bibinfo {author}
  {\bibfnamefont {P.~R.}\ \bibnamefont {Hemmer}}, \ and\ \bibinfo {author}
  {\bibfnamefont {M.}~\bibnamefont {Loncar}},\ }\href {\doibase
  10.1038/nnano.2010.6} {\bibfield  {journal} {\bibinfo  {journal} {Nature
  Nano.}\ }\textbf {\bibinfo {volume} {5}},\ \bibinfo {pages} {195} (\bibinfo
  {year} {2010})}\BibitemShut {NoStop}%
\bibitem [{\citenamefont {Ernst}\ \emph {et~al.}()\citenamefont {Ernst},
  \citenamefont {Bodenhausen},\ and\ \citenamefont {Wokaun}}]{ernst90}%
  \BibitemOpen
  \bibfield  {author} {\bibinfo {author} {\bibfnamefont {R.~R.}\ \bibnamefont
  {Ernst}}, \bibinfo {author} {\bibfnamefont {G.}~\bibnamefont {Bodenhausen}},
  \ and\ \bibinfo {author} {\bibfnamefont {A.}~\bibnamefont {Wokaun}},\
  }\href@noop {} {\bibinfo  {journal} {(International Series of Monographs on
  Chemistry, Clarendon Press, 1990)}\ }\BibitemShut {NoStop}%
\bibitem [{\citenamefont {K.Wuthrich}(2003)}]{wuthrich03jbiolnmr}%
  \BibitemOpen
\bibfield  {journal} {  }\bibfield  {author} {\bibinfo {author} {\bibnamefont
  {K.Wuthrich}},\ }\href {\doibase 10.1023/A:1024733922459} {\bibfield
  {journal} {\bibinfo  {journal} {J. Biomol. NMR}\ }\textbf {\bibinfo {volume}
  {27}},\ \bibinfo {pages} {13} (\bibinfo {year} {2003})}\BibitemShut {NoStop}%
\bibitem [{\citenamefont {Rosskopf}()}]{rosskopf16}%
  \BibitemOpen
  \bibfield  {author} {\bibinfo {author} {\bibfnamefont {T.}~\bibnamefont
  {Rosskopf}},\ }\href@noop {} {\bibinfo  {journal} {unpublished}\
  }\BibitemShut {NoStop}%
\end{thebibliography}
%merlin.mbs apsrev4-1.bst 2010-07-25 4.21a (PWD, AO, DPC) hacked
%Control: key (0)
%Control: author (8) initials jnrlst
%Control: editor formatted (1) identically to author
%Control: production of article title (-1) disabled
%Control: page (0) single
%Control: year (1) truncated
%Control: production of eprint (0) enabled
%

\end{document}